\documentclass[aip,pop,12pt,reprint]{revtex4-1}
\usepackage[dvips]{graphicx}
\usepackage{here}
\usepackage{pgfgantt}
 \usepackage{delarray}
\usepackage{amsmath}
\usepackage{amssymb}

\usepackage{ae}
\usepackage{color}
\usepackage{ifthen}
\usepackage{bm}
\usepackage{ifpdf}
\ifpdf
	\usepackage[pdftex]{hyperref}
	\hypersetup{colorlinks=true,
		pdfstartview=FitV,
		linkcolor=blue,
		citecolor=red,
		urlcolor=blue,
	}
	\DeclareGraphicsExtensions{.pdf}
\else
	\usepackage[dvips]{graphicx}
	\usepackage{epsfig}
	\DeclareGraphicsExtensions{.eps}
	\usepackage{url}
\fi

\newcommand{\beq}{\begin{equation}}
\newcommand{\eeq}{\end{equation}}

\usepackage{units}

\begin{document}

\title{Physics of the  laser-plasma interface  in the relativistic regime of interaction}
\author{B. Svedung Wettervik}
\affiliation{Department of Physics, Chalmers University of Technology, G\"{o}teborg, Sweden}

\author{M. Marklund}
\affiliation{Department of Physics, Chalmers University of Technology, G\"{o}teborg, Sweden}

\author{A. Gonoskov}
\affiliation{Department of Physics, Chalmers University of Technology, G\"{o}teborg, Sweden}
\affiliation{Institute of Applied Physics, Russian Academy of Sciences, Nizhny Novgorod 603950, Russia}
\affiliation{Lobachevsky State University of Nizhni Novgorod, Nizhny Novgorod 603950, Russia}

\begin{abstract}
The reflection of intense laser radiation from solids appears as a result of relativistic dynamics of the electrons driven by both incoming and self-generated electromagnetic fields at the periphery of the emerging dense plasma. In the case of highly-relativistic motion,   electrons tend to form a thin oscillating layer, which makes it possible to model the interaction and obtain the temporal structure of the reflected radiation. The modelling reveals the possibility and conditions for producing singularly intense and short XUV bursts of radiation, which are interesting for many applications. However, the intensity and duration of the XUV bursts, as well as the high-energy end of the harmonic spectrum, depends on the thickness of the layer  and its internal structure which are not assessed by such macroscopic modelling. Here we analyse the microscopic physics of this layer and clarify how its parameters are bound and how this controls  outlined properties of  XUV bursts.
\end{abstract}
\maketitle

\section{Introduction}

The interaction of intense radiation with overdense plasmas in the relativistic regime appears as the basic problem for many promising applications of high-intensity lasers, ranging from particle acceleration \cite{IonRev,MALKA} to the generation of high frequency radiation,\cite{HF1,HF2,HF3,HF4} with practical applications within diagnostics, probing of warm dense matter, and observing phenomena at the atto-second time-scale.\cite{APP1,APP2,APP3}

 If the transverse size of the laser pulse is much larger than the wavelength and the irradiated plasma is sufficiently dense to prevent the penetration of radiation, the reflection of an obliquely incident laser pulse can be considered as a one-dimensional problem in a reference frame moving along the plasma surface with speed $c \sin\theta$, where $c$ is the speed of light and $\theta$ is the angle of incidence\cite{bourdier.pf.1987}. In this reference frame the laser radiation impinges normally on the plasma streaming with the speed $c\sin\theta$.

The character of the laser-plasma interaction  depends crucially on the conditions of interaction. In the case of low intensity and a sharp density profile, the plasma still acts as an almost ideal mirror, but with small fractions of frontier electrons that are repeatedly thrown into the plasma resulting in heating \cite{forslund.pra.1975, brunel.prl.1987, wilks.prl.1992, price.prl.1995, kluge.prl.2011, kluge.pop.2018}. In the case of a limited density gradient, typically provided by  limited contrast of the laser pulse, the electron bunches thrown into the density ramp can excite plasma oscillations, which produce  emission of high-frequency radiation in the specular direction. This mechanism, known as coherent wake emission (CWE) \cite{quere.prl.2006}, is dominant for moderate intensities, characterized by that the field amplitude $a_0 \ll 1$, where the amplitude is given in relativistic units $mc\omega/e$ and $\omega$ is the radiation frequency, $m$ and $e$ are the electron mass and charge (absolute value).

For higher intensities, the light pressure and especially its temporal variation starts to affect the  reflection from the plasma  yielding a distinctively different mechanism for high order harmonic generation (HHG). One way of modelling this is based on the assumption that at any instance of time there exists a point where the incoming and outgoing energy fluxes are equal, and this point oscillates approaching  relativistic speed  just as an ordinary particle. Although this model is referred to as the relativistic oscillating mirror (ROM) \cite{gordienko.prl.2004}, this is   a mirror  in a limited sense: it provides phase modulations but no amplitude boosts as a real mirror would do. While this might look unnatural,  this interpretation leads to the universal law for harmonic intensity decay $I_k \sim k^{-8/3}$ where $k$ is the wavenumber, which has been observed in some simulations\cite{baeva.pre.2006} and experiments\cite{dromey.np.2006, dromey.prl.2007} (some other trends have also been discussed in the literature\cite{pirozhkov.pop.2006, boyd.prl.2008, debayle.pop.2013, boyd.pla.2016}). Note that the assumed equality of the incoming and outgoing fluxes,  known as the Leontovich boundary condition, imply that the plasma does not accumulate energy even temporarily.

\begin{figure*}
\makebox[\textwidth][c]{\includegraphics[width=1.4\textwidth]{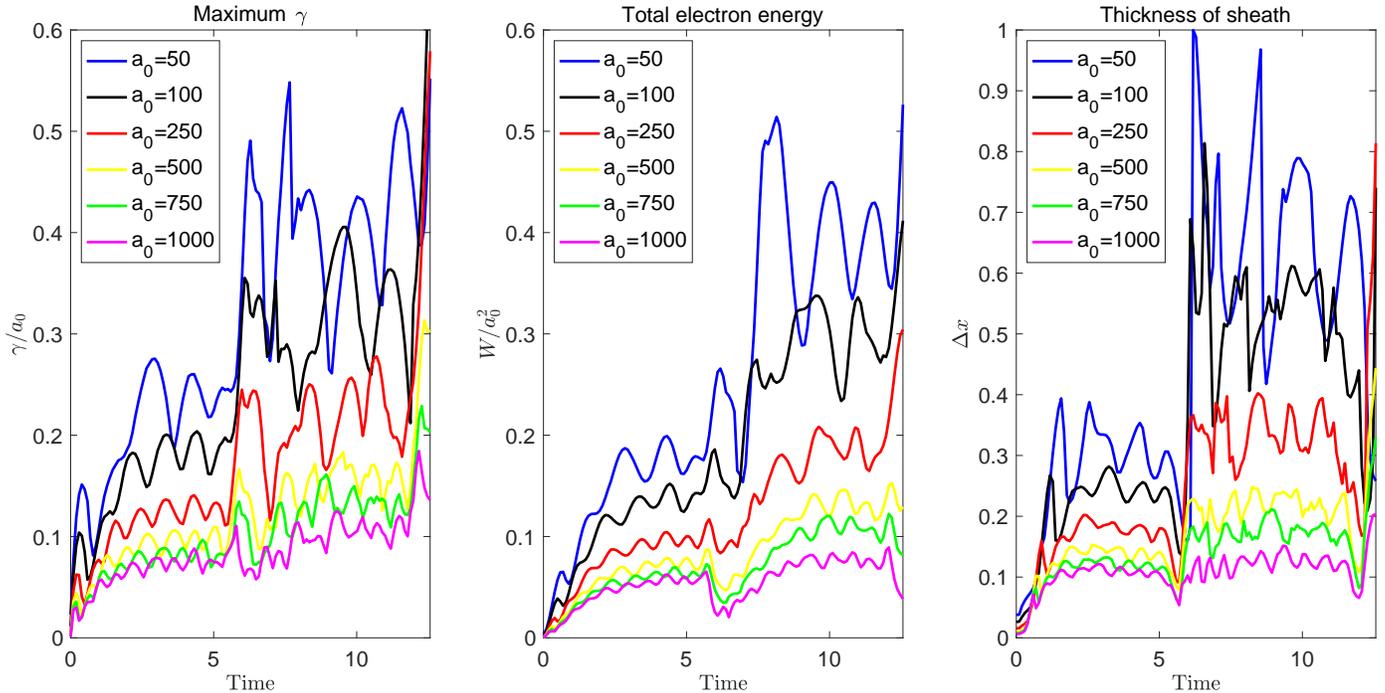}}
\caption{\label{FIG::B4}Left: Maximum $\gamma$ as a function of time for setup with $S=1/\cos^3\theta$ and angle of incidence $\theta=\pi/7$, and different $a_0$. Middle: Total energy of electrons in sheath as a function of time. Right: Thickness of sheath as a function of time. The similarity normalized quantities $\gamma/a_0$, $W/a_0^2$, $\Delta x$ decrease as $a_0\rightarrow\infty$.}
\end{figure*}

 If the intensity is even higher and/or the plasma edge is sufficiently smooth, the oscillating light pressure repeatedly causes significant back and forth shifts of the plasma front \cite{bulanov.pop.1994, lichters.pop.1996}. During these shifts a significant part of incoming radiation energy can become temporally allocated in the quasi-static field of charge separation between the shifted electrons and less mobile residual ions. In this case, the plasma acts more like a spring, repeatedly accumulating and releasing energy from the incident radiation. Simulations of this process show that  electrons tend to form a bunch and thus provide coherent synchrotron emission (CSE) \cite{quere.prl.2006}. It is notable that the bunch is compressed during both forward and backward motion due to relativistic effects and, in the case of high intensities, maintains a thickness that is much smaller than the distance over which it travels \cite{gonoskov.pop.2018}. This motivates modelling the interaction process based on treating this bunch as an infinitely thin layer that moves so that its radiation cancels out the incoming radiation into the plasma bulk. This model is referred to as the relativistic electronic spring (RES) \cite{gonoskov.pre.2011} and provides the temporal structure of the outgoing radiation for arbitrary incident radiation structure and polarization as well as for arbitrary plasma density shapes\cite{gonoskov.pop.2018}. Although challenging, some signatures of electron bunching and relativistic dynamics have already been  observed in experiments \cite{dromey.np.2012, borot.np.2012, kormin.nc.2018}.

The analysis of the RES equations indicates that at certain parameters the outgoing radiation appears in the form of singularly intense and short bursts of radiation. Simulations have showed that these bursts can have more than two orders of magnitude higher intensity than that of the incident radiation and a duration of down to a few attoseconds \cite{gonoskov.pre.2011, bashinov.epjst.2014, fuchs.epjst.2014}. One way to reach even more extreme intensities is by focusing such bursts generated from self-generated\cite{naumova.prl.2004} or manufactured \cite{gordienko.prl.2005, vincenti.nc.2018} spherical or groove-shaped\cite{gonoskov.pre.2011} plasma mirrors. Furthermore,  recently discussed applications are related to the creation of compact sources of bright XUV pulses \cite{edwards.pra.2016, lecz.josab.2018, chen.ol.2018, tang.ppcf.2018} with controllable ellipticity \cite{blanco.pop.2018} and of bright gamma rays\cite{serebyakov.pop.2015} emitted by the electrons in this regime of interaction. Since the generated XUV bursts can reach relativistic intensities for the XUV range of frequencies\cite{XUV}, they can also be used for driving wakefields in solids \cite{svedungwettervik.pop.2018, hakimi.pop.2018}. 

For the RES theory the described bursts of radiation appear as singularities and their actual peak intensity, duration and the high-energy end of the spectrum are not assessed by the theory. These characteristics are limited by the  thickness of the layer. Simulations show that, in contrast to the layer dynamics, the thickness of the layer does not follow the relativistic similarity\cite{gordienko.pop.2005} with parameter $S = n/a_0$, where $n$ is the plasma density in   units of critical density $n_{c} = m \omega^2/4\pi e^2$. This indicates that assessing the thickness requires analysis based on the first principles.

The RES model is motivated by the spread in electron velocities in the sheath being small, which is an effect of the relativistic dynamics. However, as the velocities of the electrons are close to the speed of light, small fluctuations in velocity  imply large fluctuations in the corresponding $\gamma$-factor. 
One may hence ask what happens to the similarity normalized $\gamma$-factor $\gamma/a_0$ in the high $a_0$-limit, a question that also is connected to the layer thickness $\Delta x$ and normalized energy $W/a_0^2$ of the electrons in the sheath. Figure \ref{FIG::B4} shows the similarity normalized maximum $\gamma$-factor  in the layer, the energy $W/a_0^2$, and the thickness $\Delta x$ of the sheath as a function of time, obtained from particle-in-cell (PIC) simulations,  for a range of different $a_0$. From this a decreasing trend can be observed for all three quantities as $a_0$ increases. It is not clear whether they approach a nonzero limit or converge to zero. The latter case would indicate that the efficiency of energy conversion from the laser to the electrons in the sheath becomes smaller for high $a_0$ (even if the maximum $\gamma$-factor may increase with $a_0$ in absolute numbers). The amount of energy in the compressed sheath of electrons, as well as its distribution, is relevant to address questions about the radiation spectrum for high-harmonic generation and electron heating. Resolving the $\gamma$-factor distribution for electrons in the sheath is therefore of great interest. {In previous studies, an average value of 10 has been proposed to be used as an ad hoc value for limiting the singularity of the RES equations in Ref.~\cite{gonoskov.pre.2011, gonoskov.arxiv.2018}.} Serebryakov {\it et al}\cite{serebyakov.pop.2015} have also proposed to model the average $\gamma$-factor in the sheath by solving the equations of motion for an average particle in the sheath. However, such a model is limited by the fact that particles are continuously added and removed from the sheath, leading to difficulties in connecting single-particle dynamics to that of the sheath. 

\begin{figure*}
\makebox[\textwidth][c]{\includegraphics[width=1.2\textwidth]{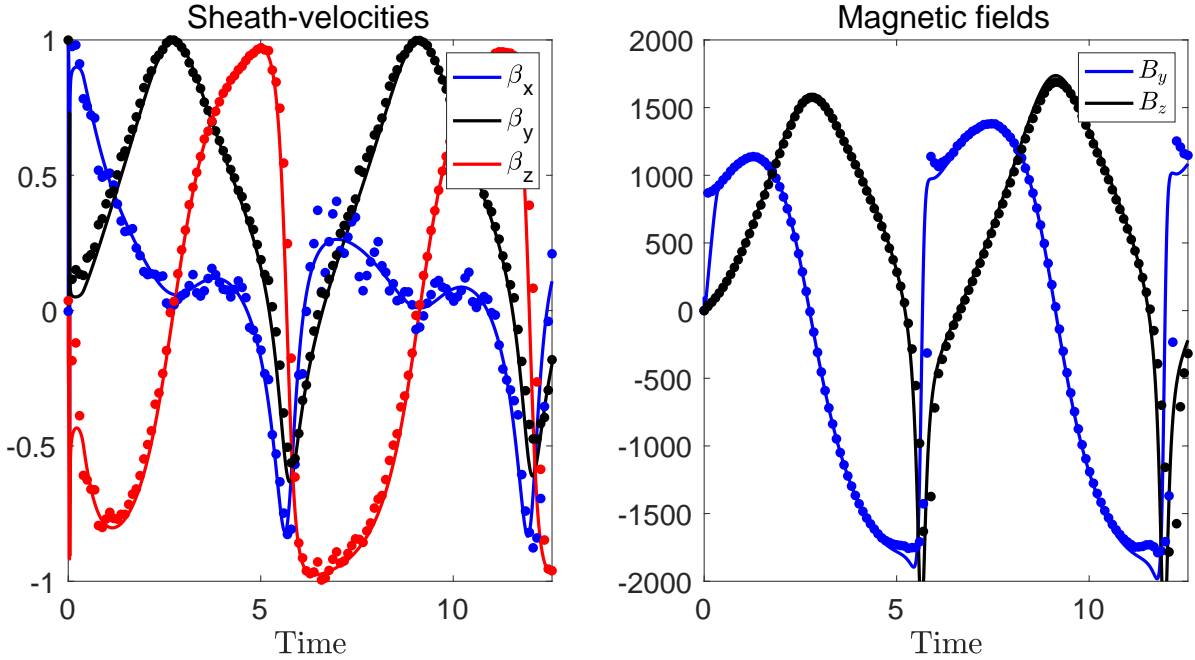}}
\caption{\label{FIG::B1}Left: Comparison of sheath-velocities from the RES-model (lines) and PIC-simulations (dots). Right: Comparison of magnetic field component at the vacuum-plasma interface from the RES-model (lines) and PIC-simulations (dots). }
\end{figure*}

In this paper, we show that the relation between the transverse momentum and vector potential can be used to express the distribution (as well as average) of  the electron $\gamma$-factors in terms of  the thickness of the  sheath and the parameters from the RES-model. However, the layer thickness and its dependency on $a_0$ still needs to be determined. One way to do this, is to integrate the rate of change of $\gamma$ for an electron at the vacuum-plasma boundary. This is highly complicated, and demands accurate models for the fields at the vacuum-plasma boundary, incorporating effects due to finite $\gamma$-factors and variations in velocities across the sheath to give a non-vanishing rate of change.  Instead,  we will here combine estimates for the layer thickness from analytical solutions with the results of particle-in-cell  simulations. The analytical estimates are based on the balance between the radiation pressure and the longitudinal electric field. Furthermore, we adress the similarity limit of the layer dynamics, as well as scaling laws for the cut-off frequency for high harmonic generation. 

The paper is structured as follows: In Sections 2 and 3, we introduce notation as well as the governing equations in the RES-model.  This is followed by Sections 4-5, where we derive the field-structure and $\gamma$-distribution inside the electron sheath. The derived expressions are compared with the results of simulations.  
Thereafter, Section 6  considers the $a_0$ dependence of the layer thickness and addresses scalings for the cut-off frequency for high-harmonics based on coherency-limits and energy conservation.  Finally, in Section 7, we summarize our findings and elaborate on possible extensions to obtain a fully analytical model for the electron and radiation spectrum, as well as discuss possible applications.

\section{Setup and units}

The RES-model can be applied for arbitrary angles of incidence, density profiles, pulse-shapes,  polarization and relativistic intensities.\cite{RES2} However, here, we consider an  incoming laser pulse of the form $\vec E(\psi)=E_{y,i}(\psi)\hat y+E_{z,i}(\psi)\hat z$, where  $\psi=\omega t-kx$ is a phase coordinate and $E_{y,i}(\psi)$, $E_{z,i}(\psi)$ are arbitrary functions of phase, interacting with a step-like plasma density profile $n(x)=n_0\Theta(x)$, where $\Theta(x)$ is a step-function and $n_0$ is the plasma density. This   can be related to the  more realistic situation with a smooth density profile by using the notion of an effective $S$-number proposed in Ref.~\cite{XUV}.  We assume that   the pulse is incident with an angle $\theta$ with respect to the $x$-axis in the plane normal to $\hat z$. By performing a Lorentz-transformation in the $\vec v=-c\sin\theta\,\hat y$ direction, the setup reduces to that of normal incidence.  However, in the boosted frame, the electrons and ions are moving with an initial velocity.

In the following, time and space are expressed in  terms of $x'=kx$ and $t'=\omega t$, where $k$ is the wave-vector of the incident radiation and $\omega$ is its frequency. Furthermore, densities  are expressed in terms of $n_c$, momentum is expressed in terms of $mc$, and fields are expressed in terms of the relativistic field $E_r=mc\omega/e$, with the relativistic amplitude defined by  $a_0=E_\text{max}/E_r$, where $E_\text{max}$ is the maximum amplitude of the incoming field. Normalizations are, unless otherwise stated,  performed with respect to the boosted frame.

\section{The Relativistic Electron Spring (RES) model -- governing equations}
\label{SEC3}
The RES-model relies on that the incident radiation will not propagate inside the plasma, i.e. only penetrates the vacuum-plasma boundary to a limited extent; eventually being cancelled by fields due to plasma-currents, in combination with $a_0>1$, which makes the electron-dynamics relativistic. Under these circumstances, the electrons form a sheath, moving with velocity $\vec\beta=(\beta_x,\beta_y,\beta_z)$, which approximately is positioned at the point of full cancellation of the incident field: $x_s$. This criteria can with respect to the boosted frame be expressed in terms of:
\begin{align*}
E_{y,i}(x_s- t)+\frac{Q}{2}\left(\sin\theta-\frac{\beta_y}{1-\beta_x}\right)=0,\\
E_{z,i}(x_s- t)-\frac{Q}{2}\frac{\beta_z}{1-\beta_x}=0,
\end{align*}
where $Q=n_0x_s$ is the total charge in the sheath. Since the dynamics is relativistic, it is assumed that the layer moves at the speed of light, i.e. $\beta_x^2+\beta_y^2+\beta_z^2=1$, resulting in three equations for the four unknowns $x_s$ and $\vec \beta_x$. The system of equations is closed by adding an equation of motion for $x_s$:
\begin{equation*}
\frac{\text{d}x_s}{\text{d}t}=\beta_x.
\end{equation*}
The fact that these equations captures the physics of interaction in the high $a_0$ limit is demonstrated in \ref{FIG::B1}.
In this Figure we compare the solution of the equations in the RES-model and PIC-simulations   for  the velocities $\vec\beta$, as well as the magnetic fields at the vacuum-plasma interface in the  particular case of $\theta=\pi/7$, $S=1/\cos^3\theta$ and  $E_{y}(\psi)=\Theta(\psi)a_0(\cos\zeta\sin\psi\,\hat y+\sin\zeta\cos\psi\,\hat z)$ with  $a_0=1000$ and $\zeta=\pi/3$, with respect to the boosted frame.
The PIC-simulations were performed using the ELMIS-code\cite{elmis}, with the spatial resolution $\Delta x=2\pi\times 2^{-14}$,  temporal resolution $\Delta t=7\times 10^{-4}$ and 25 particles per cell.

\begin{figure}
{\includegraphics[width=0.5\textwidth]{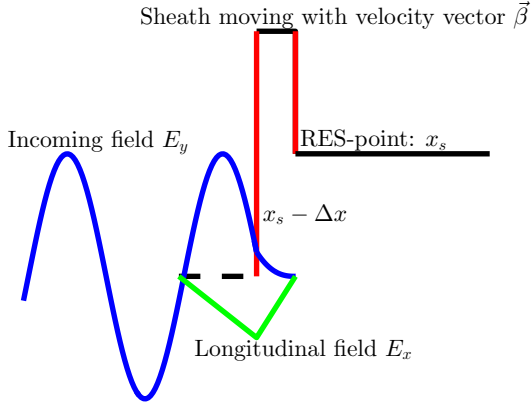}}
\caption{\label{FIG::B2}In the boosted frame, a laser pulse is incoming at a sharp plasma-boundary leading to the formation of a thin sheath with velocity $\vec \beta$. Whereas the  incoming radiation is cancelled by the $Q=n_0x_s$ electrons in the sheath, the longitudinal electric field peaks at the vacuum-plasma interface and is due to the unshielded charge  $\tilde Q=Q-n_0\Delta x$ to the left of the interface, where $\Delta x$ is the layer thickness.}
\end{figure}

\section{Fields inside the compressed electron sheath }

Figure \ref{FIG::B2} schematically illustrates a compressed sheath with velocities $\vec \beta=(\beta_x,\beta_y,\beta_z)$ and thickness $\Delta x$, resulting from the interaction between a laser and a plasma. The position $x_s$, in the RES-model, is associated with the right-most point of the sheath, whereas $x_s-\Delta x$ corresponds to the vacuum-plasma interface. The electromagnetic field at a point $x\in [x_s-\Delta x,x_s]$, i.e. inside the sheath, has four contributions: the incident field from the laser, the field from the unshielded ion current and the forward as well as backward travelling radiation from electrons inside the sheath. By using the RES-condition and summing all the contributions, a lowest order approximation to the fields at a point $x$ inside the sheath is given by:
\begin{align*}
E_x=-(Q-q),\\
E_y=(Q-q)\frac{\beta_x\beta_y}{1-\beta_x^2},\\
B_y=-(Q-q)\frac{\beta_z}{1-\beta_x^2},\\
 E_z=(Q-q)\frac{\beta_x\beta_z}{1-\beta_x^2},\\
 B_z=(Q-q)\frac{\beta_y}{1-\beta_x^2},
\end{align*}
where $q$ is the amount of electron charge between the vacuum-plasma interface and $x$.

The above expressions are used  in Section 5 to calculate the shape of the $\gamma$-factor distribution for electrons. 
However, to obtain a closed expression for the $\gamma$-distribution in the layer, it is necessary to determine either the energy in the layer, its thickness or the maximum $\gamma$-factor (the $\gamma$-factor of a particle at the vacuum plasma interface), which poses a significant challenge. In particular, with the above  field expressions both the energy-flow across the vacuum plasma boundary as well as the rate of change of $\gamma$ (i.e. $\vec\beta\cdot \vec E$) vanishes. The vanishing of energy flow across the vacuum-plasma boundary in the crude approximation of the fields does not rely on  that the sheath velocity follows that in the RES-model, but holds in the broader context of that the sheath is  described by some common velocity and moves at the speed of light. 
 To account for  energy accumulation in the sheath it is instead necessary to consider more accurate models for the field structure, which may include:
\begin{enumerate}
\item The effect of the ions situated between  $x_s-\Delta x$ and $x_s$ on $E_x$.
\item The  variation of the incoming field across the  extension of the sheath.
\item The effect of finite $\gamma$-factors.
\item  Retardation effects, both in the evaluation of the field  due to electrons and ions.
\item  Angular deviations of the particle velocities $\vec \beta(x)$ compared to the description of the sheath electrons moving with a single velocity.
\end{enumerate}
These corrections play a varying role during different parts of the interaction. Finite $\gamma$-factors on one hand directly affect the fields through the expressions for the field of an element of the sheath moving with a given velocity, but also indirectly as it determines the dynamics of the layer thickness, which broadly determines the importance of the other corrections. One may further notice that retardation plays a critical role during the emission of high-harmonics as this occurs simultaneously with the layer moving in the opposite $\hat x$-direction.

\begin{figure}
{\includegraphics[width=0.5\textwidth]{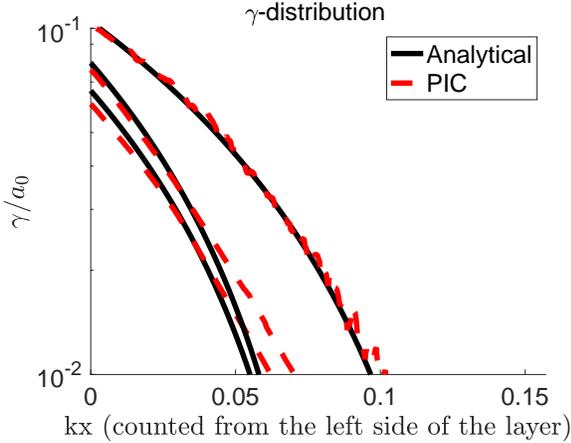}}
\caption{\label{FIG::B3}Comparison of $\gamma$-factor in the sheath calculated from PIC-simulations as well as from the $\gamma$-distribution in Section 5, with the approximation of  constant sheath-density with  thickness taken from simulations at times $t=2.0$, 4.9 and 7.9.  }
\end{figure}

\section{Estimate for $\gamma$-distribution}

Since, in the case of normal incidence, the hamiltonian of the particles does not depend on the transverse coordinates, transverse canonical  momentum is conserved. In the boosted frame, conservation of canonical momentum in the $\hat z$-direction is equivalent to:
\begin{equation*}
p_z=-A_z.
\end{equation*}
As $p_z=\gamma\beta_z$, provided that the vector-potential component $A_z$, expressed in units of $mc^2/e$,  can be calculated,  $\gamma$ is obtained by taking $\beta_z$ from the RES-model. Clearly, such expression is valid as long as the electron dynamics is relativistic  and the RES-velocity accurately describes the dynamics (which commonly is the case, except for when $\beta_z$ is close to zero).

To calculate the vector-potential, observe that  $B_y=-{\partial} A_z/{\partial}{x}$, and consequently that:
\begin{equation*}
A_z(x)=-\frac{\beta_z}{1-\beta_x^2}\int_x^{x_s} (Q-q)\text{d}x.
\end{equation*}
This integral incorporates details related to the electron-density in the layer, which from an analytical perspective cannot be known in detail (although it may be calculated from simulations). As an approximation, we assume that the density is constant across the sheath, i.e. has some value $n=Q/\Delta x$. In that case:
\begin{equation*}
A_z(x)=-\Delta x\,Q\frac{\beta_z }{1-\beta_x^2}\frac{\delta x^2}{2}
\end{equation*}
with $\delta x=(x_s-x)/\Delta x$ being the position in the sheath, normalized with the width of the sheath.

Combining the equation for the vector potential and conservation of canonical momentum gives the $\gamma$-factor distribution: 
\begin{equation*}
\gamma=\frac{Q\Delta x}{1-\beta_x^2}\frac{\delta x^2}{2}\label{EQ::gamma}
\end{equation*}
Although the analytical expression is not entirely independent from   PIC-simulations, which were used to determine the thickness of the sheath, Figure \ref{FIG::B3} indicates good consistency with the $\gamma$-factor distribution obtained from PIC-simulations.  

By integrating the $\gamma$-factor distribution, the total electron energy in the sheath can be written as:
\begin{equation*}
W(t,S,a_0)=\frac{Q^2\Delta x}{6(1-\beta_x^2)}.
\end{equation*}
Evidently, the average $\gamma$-factor  scales proportionally with $Q\Delta x$. The $S$-similarity theory implies that there is a normalized energy $W(t,S)=W(t,S,a_0)/a_0^2$, where $W(t,S)$ in the limit of high $a_0$ only depends on time and the $S$-parameter. 
In terms of $W(t,S)$, the $\gamma$-distribution takes the form:
\begin{equation*}
\gamma=3a_0W(t,S)\delta x^2/Sx_s,
\end{equation*}
which shows that the singular behaviour of $\gamma$ indicated by its dependency on $\beta_x$ is constrained by   the available energy $W(t,S)$ which is limited from above by the energy available in the laser pulse. In terms of normalized quantities, the thickness of the sheath:
\begin{equation*}
\Delta x=6(1-\beta_x^2)W(t,S)/S^{2}x_s^2,
\end{equation*}
i.e. if $W(t,S)$ converges to a limit for high $a_0$, the thickness of the sheath also converges and in particular goes to zero as the sheath moves along the axis of incidence. An estimate for the thickness in the intermediate region is given by:
\begin{equation*}
\Delta x=\frac{2}{3a_0^2W(t,S)}
\end{equation*}
where we have assumed that the velocities for the sheath can be associated with a $\gamma$, which then is equated to the maximum $\gamma$ in the sheath. However, notice that this expression is limited by the accuracy of the field-description at the point $\beta_x=-1$ and that the field at the singular point needs further consideration, which will be adressed in Section 6.

Figure \ref{FIG::B4} shows the comparison of the maximum $\gamma$, the energy of electrons in the sheath, and $\Delta x$ for   laser-plasma interaction with  $S=1/\cos^3\theta$, $\theta=\pi/7$, and  different $a_0$. Although  $\gamma$, the energy and $\Delta x$  appear to change increasingly slowly with respect to similarity normalized units as $a_0$ increases, there is no clear indication  whether they have a nonzero limit. To shed light on this, we consider the simplified problem of a circularly polarized plane wave interacting with an overdense plasma.\cite{Siminos} In this case, the balance between radiation pressure and the electrostatic force leads to a penetration depth of the vector-potential scaling as $\lambda_s\sim a_0^{-1/2}$,  i.e. which goes to zero as $a_0$ increases. If this property generalizes to the case of arbitrary interaction parameters, it would mean that $W(t,S)=0$ and consequently that energy-accumulation in the sheath is a transient phenomena, which only is significant for low to moderate $a_0$.

\begin{figure}[H]
\includegraphics[width=0.5\textwidth]{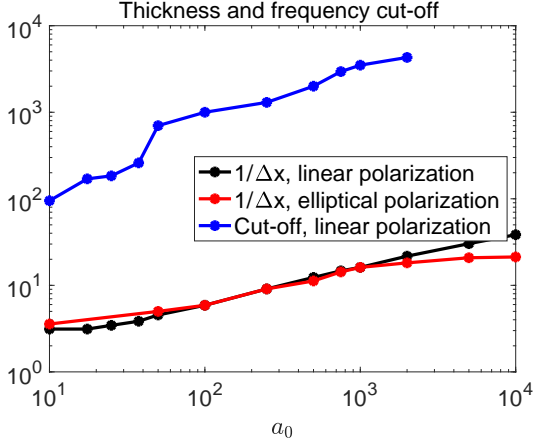}
\caption{\label{FIG::B9}The frequency cut-off for high-harmonic radiation and inverse thickness $1/\Delta x$  as a function of $a_0$  for  the interaction of a linearly polarized laser at normal incidence with an $S=1$  plasma, as well as the case with $S=1/\cos^3\theta$, $\theta=\pi/7$ and pulse shape described  in Section 3.  }
\end{figure}

\section{Simulation of layer thickness and peak field at singular point}

The existence of a limit for the layer thickness (for high $a_0$) as the layer  moves backwards  and radiates  is important for the prospects of generating coherent high harmonics of increasingly high order.  In the view of the expression for the layer thickness in Section 5, with a proportionality $\Delta x\sim (1-\beta_x ^2)$, higher $a_0$ implies a layer thickness approaching zero (as $1-\beta_x ^2\sim 1/\gamma^2$), which is a consequence of that the field approaches infinity when   $\beta_x\rightarrow-1$. However, this estimate may overestimate the decay rate by overestimating the field in the calculation of the vector potential, which is reduced due to retardation effects and potentially also affected by angular velocity spread for the electrons, which is conserved in the high $a_0$-limit.  
In Figure \ref{FIG::B9}, the  layer thickness as well as frequency cut-off for high-harmonic generation is shown as a function of $a_0$ in a case with linearly polarization, $S=1$ and $\theta=0$, as well the case described in Section \ref{SEC3}. The layer thickness is measured at the point of maximum compression (i.e. where $\beta_x=0$).  The  thickness in the two cases approximately decay as $a_0^{-0.4}$ and $a_0^{-0.3}$ respectively, which shows that the  decay rate  for the skin depth gives results that remain representative for a wider range of interaction parameters.  For the linearly polarized case, the cut-off frequency for generation of high harmonics scales as $a_0^{0.5}$, which is fairly consistent with the limit on coherency implied by $\sim 1/\Delta x$ and the scaling of the layer thickness.

An implication from the study of the scaling of the thickness is that the energy in the sheath scales as $W(t,S,a_0)\sim a_0^{2-\alpha}$, where $\alpha \approx 0.5$ is such that $\Delta x\sim a_0^{-\alpha}$.  Assuming the atto-second burst generated from the interaction can be described by an amplitude $B$ and typical wavelength $L$: $LB^2 \sim a_0^{2-\alpha}$, which provides a different route to the scaling of the frequency for high harmonic generation. To find $L$, it is necessary to estimate the amplitude of the atto-second burst. 
Here, we take into account the velocity spread through a delay $\tilde t(q)$ such that $\beta_x(t,q)=\rho\beta_x(t-\tilde t(q))$, where $\rho=\sqrt{1-1/\gamma^2}$ and $\vec \beta (t)$ corresponds to a motion at the speed of light. In the vicinity of an electron with $\beta_x(t,q)=-\rho$ at   time  $t=t_e-\tilde t(q)$:
\begin{equation*}
\beta_x(t,q)=\rho\left(-1+\frac{1}{2}\frac{\partial^2\beta_x}{\partial t^2}(\delta t-k\delta x)^2\right)
\end{equation*}
where $k=n\partial \tilde t/\partial q$ and $\delta t$, $\delta x$ are the deviations from the position and time   $t=t_e-\tilde t(q)$. It holds that:
\begin{equation*}
\beta_y(t,q)\sim  \sqrt{\frac{\partial^2\beta_x}{\partial t^2}}(\delta t-k\delta x),
\end{equation*}
and consequently:
\begin{equation*}
\text{d}B_z\sim  \frac{\sqrt{\frac{\partial^2\beta_x}{\partial t^2}}(\delta t-k\delta x)}{1-\rho+\frac{1}{2}\frac{\partial^2\beta_x}{\partial t^2}(\delta t-k\delta x)^2}\text{d}q.
\end{equation*}
Taking into account retardation amounts to setting $\delta x=-\delta t$ and the total field can be obtained by integration of all contributions. However,   the integrand is anti-symmetric in $ \delta x$ and nonzero values of the field  are hence a consequence of the layer position for the zero-crossing of the transverse velocity (with the correct phase) as well as variations of $\gamma$ across the layer. An upper estimate for the field that can be obtained by only taking into account the constructive field contributions:
\begin{equation*}
B_z \sim \frac{n\log\gamma}{(1+k)\sqrt{\frac{\partial^2\beta_x}{\partial t^2}}}.
\end{equation*}
This expression is proportional to $n$, which as $n\sim Q/\Delta x \sim a_0^{1+\alpha}$ shows that the field grows no faster than $a_0^{1+\alpha}\log a_0$. Combining this with the estimate for energy: $L\sim a_0^{-3\alpha}/(\log a_0)^2$, which translates into a frequency scaling $\tilde\omega/\omega\sim a_0^{3\alpha}(\log a_0)^2$, where $\tilde \omega$ is a typical frequency for the attosecond burst. Energy constraints hence allow a faster increase of the frequency with $a_0$ than the scaling of the layer thickness. The frequency of high harmonics may hence be anticipated to scale at this rate until it reaches high enough values for coherency to set limits, then following the slower scaling of $1/\Delta x$.

\section{Conclusions}

Understanding the properties, e.g. frequency range, amplitude and duration of high harmonics generated from the interaction of lasers with moderately overdense plasma, has been of great experimental as well as theoretical interest. The RES-model has  in the past been shown to  model the layer dynamics for laser-plasma interaction in the near-critical regime, which is relevant to the generation of high harmonic radiation.  However, to understand the details of the radiation spectrum, it is necessary to assess the electron spectrum and properties of the micro-dynamics for electrons in the sheath.

In this paper, we connected the  $\gamma$-factor inside the electron layer to  RES-parameters and the thickness of the electron sheath or, equivalently, in terms of the total energy of the  sheath. However, it was observed that the energy-flow across the vacuum-plasma boundary vanished, unless field-effects of the order of the thickness of the layer were included. Such field effects include variations of the velocities across the sheath, retardation effects, as well as effects due to finite $\gamma$-factors for the electrons.  

Finally, guided by analytical estimates and simulations, we found that the layer thickness $\Delta x$ scales as  $a_0^{-\alpha}$, where $\alpha\sim 0.5$. Based on these observations, two scalings for the cut-off frequency for high harmonics could be indicated. On the one hand from limits due to energy constraints and on the other hand from  incoherency ($\sim 1/\Delta x$). Such scalings are consistent with a faster growth of the cut-off frequency for small $a_0$ than at higher $a_0$. To improve the accuracy of the analysis of the radiation generation it is suggested to focus on a more in detail understanding of the scaling of the layer thickness with the relativistic amplitude  as well as effects of the micro-dynamics of the electron-sheath during emission of high harmonics.

\section{Acknowledgements}

This research was supported by the Knut \& Alice Wallenberg Foundation Grant Plasma based compact ion sources, and the Swedish Research Council (grant 2016-03329 and 2017-05148). The simulations were performed on resources provided by the Swedish National Infrastructure for Computing (SNIC) at High Performance Computing Center North (HPC2N). 

\bibliography{paper}

\end{document}